\documentclass[12pt,preprint]{aastex}       
\usepackage{psfig}

\begin{document}
 
\title{Coupling Between Periodic and Aperiodic Variability in 
SAX J1808.4-3658}

\shorttitle{Harmonic Coupling of the Red Noise in SAX J1808.4-3658}
\shortauthors{Menna et al.} 

\author{M. T. Menna}
\affil{Osservatorio Astronomico di Roma, via Frascati 33, I--00040 
Monteporzio Catone (Roma), Italy.}
\email{menna@coma.mporzio.astro.it}

\author{L. Burderi}
\affil{Osservatorio Astronomico di Roma, via Frascati 33, I--00040 
Monteporzio Catone (Roma), Italy.}

\author{L. Stella}
\affil{Osservatorio Astronomico di Roma, via Frascati 33, I--00040 
Monteporzio Catone (Roma), Italy.}

\author{N. Robba} \affil{Dipartimento di Scienze Fisiche ed
Astronomiche, Universit\`a di Palermo, via Archirafi 36, I--90123
Palermo, Italy.}

\author{M. van der Klis} \affil{Astronomical Institute Anton
Pannekoek, University of Amsterdam, Kruislaan 403, 1098 SJ
Amsterdam, The Netherlands}

\begin{abstract}
  We detect a significant broadening in the wings of the $401$ Hz peak
  in the power spectrum of the accreting millisecond binary pulsar SAX
  J1808.4-3658. This feature is consistent with the convolution of the
  red noise present in the power spectrum with the harmonic line. We
  conclude that the flux modulated by the spin period shows aperiodic
  variability similar to the red noise in the overall flux, suggesting
  such variability also originates at the magnetic caps close to the
  neutron star surface. This is analogous to the results found in some
  longer period, higher magnetic field pulsators in high mass X-ray
  binaries.
\end{abstract}   


\section{Introduction}
The power density spectra of low mass X-ray binaries show broad band
noise components, produced by aperiodic variability, the sum of which
typically rises towards lower frequencies (sometimes flattening below
a frequency $\nu_{\rm knee}$) and is often termed red noise.  Other
power spectrum features are often present, including relatively broad
peaks, quasi periodic oscillations, bumps, wiggles and low frequencies
excesses (see {\it e.g.} Wijnands \& van der Klis 1999 for a review).
In many respects these noise components are similar to those observed
in high mass X-ray binaries, containing X-ray pulsators and black hole
candidates (see {\it e.g.} Belloni \& Hasinger 1990).  The continuum
component of the red noise is usually modeled by broken power laws
and/or Lorentzians.

In power spectra of X-ray pulsators a number of harmonic lines arising
from the modulation of the flux due to the neutron star spin is also
present.  In this case, information on the regions giving rise to the
aperiodic variability can be extracted from the study of the power
spectra (Burderi {\it et al.}, 1997, hereafter B97; Lazzati and
Stella, 1997, hereafter LS97). Indeed, if all or part of the aperiodic
variability is produced by matter funneled by the magnetic field lines
and accreting onto the magnetic polar caps of the neutron star, a
modulation of that part of the aperiodic signal at the spin period
would be expected (harmonic coupling). In the power spectra this would
appear as a broadening of the wings of the harmonic lines due the
convolution of each harmonic line with the red noise (Burderi, Robba
\& Cusumano 1993) that produces a rescaled version of the red noise
placed at both sides of each harmonic peak.  Alternatively, if none of
the aperiodic variability is generated in the funneled matter, the
periodic and red noise features are expected to be independent of each
other and there should be no broadening at the base of the harmonic
lines.

A significant broadening in the wings of the harmonic lines was
detected in the power spectra of some X-ray pulsators (B97; LS97),
indicating that aperiodic variability is modulated at the spin period
of the neutron star, {\it i.e.}  produced at the neutron star magnetic
polar caps. More recently, a coupling between the period and aperiodic
variability of Hercules X-1 has been reported (Moon \& Eikenberry
2001).

In 1998 the first millisecond accretion powered X-ray pulsar was
discovered with the Rossi X-ray Timing Explorer (RXTE) by Wijnands \&
van der Klis (1998a). The pulsation frequency was $\sim 401$ Hz and
showed a periodic modulation at $2.01$ hours testifying to the
presence of a low mass companion orbiting the neutron star with a mass
function of the secondary star of $3.8 \times 10^{-5} M_{\odot}$
(Chakrabarty \& Morgan 1998). The source was identified with SAX
J1808.4-3658, a transient first detected with Beppo SAX in 1996 which
displayed type I X-ray bursts and was classified as a low mass X-ray
binary at an estimated distance of 4 kpc (in 't Zand et al. 1998).
This distance has been recently updated to 2.5 kpc (in 't Zand et al.
2001). The coherent pulsation was detected during the 1998 outburst
when the source was repeatedly observed with RXTE from the $11^{th}$
of April to the $6^{th}$ of May.  The source reached a peak luminosity
of $ 3 \times 10^{36}$ erg s$^{-1} $ on the $13^{th}$ of April and
then faded gradually below $10^{35} $ erg s$^{-1} $ by the $6^{th}$ of
May.  Throughout this period the X-ray spectrum remained stable,
featuring a power law shape with a photon index $ \sim 2 $ in the
15-100 keV band and a high energy cut-off above about 100 keV, similar
to the X-ray spectra of other type I bursters in their low state
(Gilfanov et al.  1998; Heindl \& Smith 1998).  Wijnands \& van der
Klis (1998b) studied the rapid aperiodic variability and revealed the
presence of a red noise component in the power spectra flattening
below a knee frequecy $\nu_{knee} \sim 1$ Hz. The periodic modulation
was nearly sinusoidal in shape and the power spectra displayed a very
pronounced coherent peak at $\sim 401 $ Hz.

The possibility of detecting a modulation at the spin frequency in the
aperiodic variability depends crucially on the spin frequency: if this
is too low, {\it i.e.} $\nu_{\rm spin} \ll \nu_{\rm knee}$ the
broadening of the harmonic line's wings is so wide that it blends in
the red noise itself (B97; LS97).  However, for SAX J1808.4-3658 $
\nu_{\rm spin} \gg \nu_{\rm knee} $, which means that the typical
duration of the aperiodic variations is many spin cycles, making this
source a particularly well suited candidate to search for the presence
of a coupling between this red noise component and the periodic
variability.

\section{The model}
\label{sec:model} 

Analytic derivations of the power spectrum of a signal consisting of
aperiodic variability, a periodic modulation and a coupling of the two
was described by B97 and LS97. Our aim is to use an analytic function
to fit the power spectrum of the SAX J1808.4-3658 to derive the entity
(if present) of the coupling. In line with B97, we adopt randomly
occurring shots as a convenient description of the aperiodic
variability giving rise to the red noise (shot noise model).  In this
model the aperiodic variability is produced by a random superposition
(in the light curve) of shots. These instabilities might originate in
the accretion flow onto the compact object. We note that we can apply
this model without loss of generality, as the coupling effect is
independent of the mathematical model adopted to describe the red
noise. In the following we briefly summarize some useful definitions
and results from B97.

Our shot noise process ${\cal N}_{\rm sn}(t)$ is obtained from the
random superposition of identical shots occurring at a constant mean
rate $\lambda$: the shot magnitude is $S = \int h(t) dt$, where $h(t)$
is the intensity profile of a shot. This means that the average
intensity of the shot component is $\bar I_{\rm sn} = \lambda S $.  We
indicate the power spectrum of the individual shot as $|H(\nu)|^2$ .

A signal $I(t)$, as seen by a distant observer, emitted by a rotating
neutron star of spin period $P$ is, in general, composed by: \\ a) a
(constant) background component $I_{\rm bck}$; \\ b) an emission
(diffuse) arising from matter not funneled on the neutron star
magnetic caps $I_{\rm DF}(t)$ and therefore not affected by the
periodic modulation induced by the neutron star rotation; \\ c) a
modulated (localized) component arising from matter accreting onto the
magnetic poles $I_{\rm LC}(t)$.  The periodic modulation function
$M(t)$, representing the geometric lighthouse effect produced by the
neutron star rotation, is a series of identical pulse profiles of
period $P=1/\nu_0$.  Expanded in Fourier series $ M(t) = C + \sum_{k}
c_k \cos( {2\pi \, k \, \nu_0} t + \phi_k) $. In the case of SAX
J1808.4-3658 the pulse profile is almost sinusoidal and we can
approximate $M(t) \simeq C + c_1 \cos( {2\pi \, \nu_0} t) $. We
require $0 \le M(t) \le 1$, this poses a constraint on $C$ and $c_1$,
namely, $C=1-c_1$. Thus we have
$$ 
I(t) = I_{\rm bck} + I_{\rm DF}(t) + I_{\rm LC}(t) \times M(t).  
$$
Let us assume that the diffuse and localized emission are each
composed of a shot noise component and a constant component: we assume
that the shot magnitude $S$ and profile $h(t)$ is the same for both
the shots arising in the localized and in the diffuse regions, while
the shot rates are respectively $\lambda_{\rm LC}$ and $\lambda_{\rm
  DF}$.  $K_{\rm DF}$ and $K_{\rm LC}$ are respectively the constant
emission of the diffuse and localized components of the signal.  Thus
we have:
$$
I_{\rm DF}(t) = K_{\rm DF} + {\cal N}_{\rm sn} (t)_{\rm DF}
$$
$$
I_{\rm LC}(t) = K_{\rm LC} + {\cal N}_{{\rm sn}} (t)_{\rm LC}
$$
and for the mean total intensity:
$$
\bar I = I_{\rm bck} + ( K_{\rm DF} + \lambda_{\rm DF} S )  + ( K_{\rm LC} + \lambda_{\rm LC} S )  \times (1-c_1) 
$$

The power spectra of the signal described above, taking into
account the finite duration $T$ of the light curves and adopting the
normalization of Leahy (1983) is:
\begin{equation} 
P(\nu) = P_{\rm 0} + P_{ \rm RN} + P_{ \rm HL} + P_{ \rm CPL} + P_{ \rm WN} 
\label{eq:PSD} 
\end{equation} 
The first term is the power at zero frequency and is only dependent on
the total intensity of the signal.
\begin{equation} 
P_{\rm 0} =  N  {\bar I}^2 T W^2_T(\nu)
\end{equation} 
where $N = 2/\bar I$ is the normalization constant according to Leahy
and where $W^2_T(\nu) = \sin^2(\pi \nu T) / (\pi \nu T)^2 $ is the
window function. We note that $P_{\rm 0} =0 \; \forall \nu_k = k/T$
where the Fourier frequencies are computed at $k=1,...,(T / 2 \Delta
t) - 1 $.  For this reason $P_0$ has not been included in the fit
discussed in \S \ref{sec:fit}.
The red noise (RN) term is
\begin{equation} 
P_{ \rm RN} = N \times ( \lambda_{\rm DF} + (1-c_1)^2  \lambda_{\rm LC} )  S^2  f_{\rm RN}(\nu) 
\label{eq:RN} 
\end{equation} 
where, in the simplest case of a shot noise process obtained from the
random superposition of identical shots, $f_{\rm RN}(\nu) = |H(\nu)|^2
/ {\rm MAX} \{ |H(\nu)|^2\} $, in a more general case $f_{\rm
  RN}(\nu)$ is a non negative function normalized to $1$ that
describes the shape of the red noise. The harmonic line term is given by:
\begin{equation} 
P_{ \rm HL} = N \times (c_1/2)^2 \times ( K_{\rm LC} + \lambda_{\rm LC}  S)^2   \times  T W^2_T(\nu-\nu_0) 
\label{eq:HL} 
\end{equation} 
where, because of the almost sinusoidal shape of the SAX J1808.4-3658
pulse profile, we have taken into account only the fundamental
frequence.  The coupling term is:
\begin{equation} 
P_{ \rm CPL }= N \times (c_1/2)^2  \times \lambda_{\rm LC}  S^2  f_{\rm RN}(\nu-\nu_0)  
\label{eq:CPL} 
\end{equation}
It consists in a rescaled versions of the red noise placed at the
harmonic frequencies, the scaling factor is $r= P_{ \rm CPL }/P_{ \rm
  RN} = {1 \over 4} \left ( {c_1 \over (1-c_1)} \right )^2 \left (1 +
  {\lambda_{\rm DF} \over (1-c_1)^2 \lambda_{\rm LC} } \right )^{-1} $.  The
white noise component $P_{ \rm WN}$, induced by counting statistics,
has, adopting the normalization of Leahy, an expected value of $2$.
The actual value is usually (slightly) less than $2$ because of
detector dead-time effects ({\it e. g.} van der Klis 1989), thus its
value will be determined by the fit of the power spectra, as described in \S
\ref{sec:fit}.

We define the coupling fraction $ \eta = \lambda_{\rm LC} /(
\lambda_{\rm LC} + \lambda_{\rm DF}) $ as the ratio of the arrival
rates of the localized shots over the total shot rate.  Furthermore we
define the pulsed fraction ${\cal P_{\it f}} = { ( M(t)_{\rm MAX} -
  M(t)_{\rm MIN} ) / M(t)_{\rm MAX} } = 2 c_1$, where
$M(t)_{\rm MAX}$ and $M(t)_{\rm MIN}$ are respectively the maximum and
the minumum of the modulating function over one period. With these
definitions $ 0 \leq {\cal P_{\it f}} \leq 1 $ and we have $c_1 / (1-c_1) = {\cal
  P_{\it f}} / (2 - {\cal P_{\it f}})$. Moreover, as $ M(t) \geq 0 $
always, we have $ 1/2 \leq (1-c_1) \leq 1 $ and $ 1 \leq 1/(1-c_1)^2 \leq 4$.
Therefore the following relation must hold:
\begin{equation} 
{4 \sqrt{r}  \over \left[ 1+ \left( { 1 - \eta \over \eta} \right) \right]^{-1/2} + 2 \sqrt{r} } \le {\cal P_{\it f}} \le {4 \sqrt{r}  \over \left[ 1+ 4\left( { 1 - \eta \over \eta} \right) \right]^{-1/2} + 2 \sqrt{r} }
\label{eq:pf} 
\end{equation}
In the following section we will determine $r$ from the fit of the
power spectra and with relation (\ref{eq:pf}) we will constrain the
values of the pulsed fraction in relation to the fraction of coupled
shots.

Moreover we will be able to derive a relation between the diffuse
component of the signal, as defined above, and the pulsed fraction.
Averaging the signal only over the shot noise process we define:
\begin{equation}
\langle I(t) \rangle=  I_{\rm bck} + ( K_{\rm DF} + \lambda_{\rm DF} S )  + ( K_{\rm LC} + \lambda_{\rm LC} S )  \times (C + c_1 \cos( {2\pi \, \nu_0} t))  
\end{equation}
thus 
$$
\langle I(t) \rangle_{\rm MAX} =  I_{\rm bck} + ( K_{\rm DF} + \lambda_{\rm DF} S )  + ( K_{\rm LC} + \lambda_{\rm LC} S ) 
$$
$$
\langle I(t) \rangle_{\rm MIN} =  I_{\rm bck} + ( K_{\rm DF} + \lambda_{\rm DF} S )  + ( K_{\rm LC} + \lambda_{\rm LC} S ) \times (1 - 2 c_1)
$$
We can therefore define the ``observed'' pulsed fraction  
$${\cal P_{{\it f}\,{\rm obs}}} = { ( \langle I(t)\rangle_{\rm MAX} - \langle I(t) \rangle_{\rm MIN} ) / ( \langle I(t)\rangle_{\rm MAX} - I_{\rm bck} ) } 
$$ with a little algebra we can derive ${\cal P_{{\it f}\,{\rm obs}}}
= 2c_1 / \left( { K_{\rm DF} + \lambda_{\rm DF}S \over K_{\rm LC} +
\lambda_{\rm LC}S } +1 \right)$ = $ 2c_1 / \left( { \langle I_{\rm DF} \rangle \over
\langle I_{\rm LC} \rangle } +1 \right) $. Recollecting that ${\cal P_{\it
f}} = 2c_1 $ we can derive $ { \langle I_{\rm DF} \rangle \over
\langle I_{\rm LC} \rangle } = { \cal P_{\it f} \over {\cal P_{{\it
f}\,{\rm obs}}} } -1$. We can now define the diffuse fraction as the
ratio of the diffuse component over the sum of the diffuse and
localized components as $ \xi = { \langle I_{\rm DF} \rangle \over
\langle I_{\rm DF} \rangle + \langle I_{\rm LC} \rangle }$. Using the
relations above we have $ \xi= 1- {{\cal P_{{\it f}\,{\rm obs}}}
\over {\cal P_{{\it f}}} } $. We will see in the next section how the
constraints on the pulsed fraction will have a bearing also on the
fraction of the diffuse component.

\section{Data analysis and results}
\label{sec:fit}
We used the observations of SAX J1808.4-3658 performed by the Rossi
X-Ray Timing Explorer (RXTE) during the April-May 1998 outburst (see
Table 1). In particular we examined the data from the Proportional
Counter Array (PCA) (Jahoda et al. 1996) which is composed of a set of
five xenon proportional counters sensitive to X-rays in the 2 - 60 keV
range with a total area of $ \sim 7000 cm^2 $ and a $ 1^{\circ} $ FWHM
field of view. We used data collected with $\sim 122 \mu{\rm s}$ time
resolution and 64 PHA channels which were available for all
observations (except for the April 13 observation which was excluded
from our analysis).

We first corrected photons arrival times for satellite motion with
respect to the solar system barycenter using the BeppoSax position for the
source (in 't Zand et al. 1998) $R.A. = 18h 08m 29s$, $Dec =
-36^{\circ}$ $58.6' (J2000)$.  We then corrected for the orbital
motion of the neutron star by using the orbital parameters derived by
Chakrabarty \& Morgan (1998). For our analysis we rebinned the data
into bins of $\Delta t = 488.28125 \mu{\rm s}$ (corresponding to a
Nyquist frequency of $1024$ Hz) and we used the events from all 64 PHA
channels. The resulting lightcurves were used for the subsequent
analysis.

For each observation, we calculated consecutive FFTs over time
intervals of $T = 64 {\rm s}$ corresponding to power spectra with a frequency
range of 0.015625-1024 Hz. These power spectra were then averaged to produce a
single power spectra for each observation, 16 in all.

The source flux decreased across the observations by a factor of $
\sim 100$ (Gilfanov et al.  1998 ).  The aperiodic variability was
determined to change with flux level (Wijnands \& van der Klis,
1998b). On the other hand, the simple model of the aperiodic
variability described in the previous section assumes that the
characteristics of the signal do not vary with time.  Thus in our
analysis we selected and analyzed only the observations in which the
source count rate was highest (between $\sim 800$ and $\sim 500
cts/s$).  In particular we selected the 6 consecutive observations
between April 11 and 20.  Examining the power spectra of each
observation we consistently find an almost constant value for the red
noise intensity and shape. 

The shape of the red noise in SAX J1808.4-3658 was found to be reasonably
well described by the sum of three empirical functions:
$$
f_{\rm RN}(\nu) = \sum_{i=1}^{3} p_i \, f_i(\nu) 
$$
where: $ f_i(\nu) = [1+(2\pi \tau_i \nu )^{\alpha_{i}}]^{-1} $ and
$p_i$ indicates the normalization of each of them.

In order to fit the spectrum efficiently, we logarithmically rebinned
the spectrum so that the number $n_i$ of original bins averaged in the
$i^{\rm th}$ new bin follows $n_i={\rm int}(1.02^i)$, where ${\rm
  int}(x)$ is the integer part of $x$. The region around the coherent
modulation peak ($400.8< \nu < 401.2$) was left at the original Fourier
resolution, while the intervals $352.37<\nu\le400.8$ and $401.2 \le
\nu < 449.64$, in which the peak broadening would be more important,
were (uniformly) rebinned by a factor of $31$ (see Fig.  1). The
region around the second harmonic of the modulation was excluded (see
above).
We used a fitting function :
$$
F(\nu) = A_1 f(\nu) +  A_2 g(\nu - \nu_0) +  A_3 f(\nu - \nu_0) + A_4
$$ where $g(\nu) = W^2_T(\nu) $ and $f(\nu) = p_1 f_1(\nu) + p_2
f_2(\nu) + p_3 f_3(\nu) $ as defined above. Thus $A_1 f(\nu) = P_{\rm
RN}$, $ A_2 g(\nu - \nu_0) = P_{\rm HL}$, $A_3 f(\nu - \nu_0) = P_{\rm
CPL}$ and $A_4 = P_{\rm WN}$.  

In order to check the robusteness of our fit we compared different
fitting procedures. a) In the first place we have tried a global fit,
that is we simultaneously fit the whole range leaving all the
parameters of the fit free to vary. Then, keeping fixed the parameters
that describe the three empirical functions $f_i$, namely $p_i$,
$\alpha_i$, $\tau_i$, we compare the fit on the whole range leaving
$A_1$, $A_2$ and $A_3$ free and leaving just $A_1$ and $A_2$ free and
forcing $A_3=0$, that is imposing the hypothesis of null coupling. We
obtain two values of the $\chi^2$ with $\Delta \chi^2 = \chi^2_1 -
\chi^2_2 = 19$. b) In second instance we repeated the procedure we
compare the fit before and after freezing $A_3=0$ only in the region
near the armonic line, beetween $396$ and $406$ Hz, the difference in
the $\chi^2$ we obtain is $\Delta \chi^2 = 21$. c) Last we wanted to
check if the effect was still present freezing also $A_2$, measuring
the height of the harmonic line, thus we compared the fit in the same
region of case b) and found that the $\Delta \chi^2=21$. We
thus conclude that the region that affects the fit is only the narrow
region around $401$ Hz and a significant coupling is present
independently of the detailed fitting procedure. The values determined
for case b) of the $f_i$ and $A_1$, $A_2$, $A_3$ parameters are listed
in Tab. 2 along with their errors. From the best fit parameters $A_3$
and $A_1$ we derive $r= P_{\rm CPL}/P_{\rm RN} = 0.005$. An F--test
calculates the probability that the $\Delta \chi^2$ variation is due
to a chance improvement $P=8.38 \times10^{-6} $.

With the derived value of $r$, we plot in figure (\ref{fig:pf}) the
two constraints defined by equation (\ref{eq:pf}) in the $\eta$ -- $
{\cal P_{\it f}}$ plane. The area between the curves defines the
allowed region for the parameters. It can be seen that for any allowed
value of the pulsed fraction a fraction of shots between $0.8-100\%$
must be modulated at the spin period (coupled). Moreover, even in the
most favourable case of $100\%$ localized shots, the modulation has to
be $> 17\%$ in order to account for the broadening detected. It can be
noted that, in our discussion, we have assumed an identical amplitude
for the localized and diffuse shots, that is, we have assumed that the
blobs originating in the disk survive intact to the polar caps. It
might be argued that the magnetic field just scoops off a fraction of
each blob. We examined, for example, a case in which just $10\%$ of
each blob is threaded to the neutron star surface. In this scenario we
obtain that nothing much changes in the favourable case of total
coupling ($100\%$), whilst for the maximum value of the pulsed
fraction ($100\%$), a minimum fraction of shots greater than $45\%$
must be modulated at the spin period. That is, if just a part of each
blob is scooped off, then more blobs must be funneled to observe the
same effect.

A different constraint can be derived comparing the ``observed''
pulsed fraction value $ {\cal P_{{\it f}\,{\rm obs}}}$ with its
theoretical value $ {\cal P_{\it f}}$ as derived from eq. 6 above. To
compute the value of the ``observed'' pulsed fraction we have folded the
data corrected for the orbital motion modulo the spin period and
obtained a pulse profile. Adopting the background value for this
observation of $130$ c/s (Wijnands \& van der Klis 1998a ) we found
${\cal P_{{\it f}\;{\rm obs}}} \sim 12\%$. We can note from fig. (2)
that the pulsed fraction $0.17 \leq {\cal P_{\it f}} \leq 1 $ and,
recollecting the definition of the fraction of diffuse component $\xi
= 1 - {{\cal P_{{\it f}\,{\rm obs}}} \over {\cal P_{{\it f}}} }$ we
can pose a constraint on $\xi$, 
$ 1- { {\cal P_{{\it f}\,{\rm obs}}} \over 1 } \leq \xi \leq 1- { {\cal P_{{\it f}\,{\rm obs}}} \over
{\cal P_{{\it f}\,{\rm min} }}} $, and so $0.294 \leq \xi \leq
0.88$. We can now plot in fig (3) the diffuse fraction $\xi$, versus
the fraction of coupled shots $\eta$.

\section{Summary and Conclusions}

We have shown the presence of a broadening at the base of the harmonic
line. This can be interpreted as the result of a coupling between a
fraction (between $0.8\%$ and $100\%$) of the aperiodic and periodic
variability in SAX J1808.4-3658. Moreover we have demonstrated that a
fraction between $\sim 30\%$ and $\sim 90\%$ of the total emission
arises from a component that is not modulated by the lighthouse effect
(diffuse component).  

Some more stringent constraints can be derived if we assume that the
diffuse component originates from an accretion disk and the localized
component derives from accretion onto the neutron star surface.  As
the spin period of SAX J1808.4-3658 is known, we can compute the
corotation radius for this system {\it i.e.}  the radius at which the
speed of a particle rigidly rotating with the neutron star equals the
local Keplerian value, $R_{\rm co} = 1.5 \times 10^{6} m^{1/3}
P^{2/3}_{-3}$ cm, where $m$ is the NS mass in units of solar masses
and $P_{-3}$ is the NS spin in milliseconds. In our case we get
$R_{\rm co} = 30.8$ km assuming a mass of $1.4 M_{\odot}$. For the
accretion not to be centrifugally inhibited the inner rim of the disc
must be located inside the corotation radius. Therefore we have
$R_{\rm disc} \leq R_{\rm co}$. In the standard scenario of disk
accretion onto magnetized neutron stars, the accreting matter forms a
disk whose inner radius is truncated at the magnetosphere by the
interaction of the accretion flow with the magnetic field of the NS.
In this case the magnetospheric radius $r_{\rm m}$ is a fraction $\phi
\la 1$ (an expression for $\phi$ can be found in Burderi et
al. 1998\footnote{$\phi = 0.21 \alpha^{4/15} n_{0.615}^{8/27}
m^{-142/945} [(L_{37}/\epsilon)^{8/7} R_6^{8/7}
\mu_{26}^{5/7}]^{4/135}$, see the text for the definition of the
symbols.}; for $L \sim 10^{36}$ ergs/s we get $\phi \sim 0.3$) of the
Alfv\'en radius $R_{\rm A}$ defined as the radius at which the energy
density of the (assumed dipolar) NS magnetic field equals the kinetic
energy density of the spherically accreting (free falling) matter:
\begin{equation}
R_{\rm A} = 2.23 \times 10^6  R_6^{-2/7}  m^{1/7}
\mu_{26}^{4/7} \epsilon^{2/7} L_{37}^{-2/7} \;{\rm cm}
\end{equation}
(see, e.g., Hayakawa 1985), where $R_6$ is the NS radius, $R_{\rm
NS}$, in units of $10^6$ cm, $m$ is the NS mass in solar masses,
$\mu_{26}$ is the NS magnetic moment in units of $10^{26}$ G cm$^3$,
$\epsilon$ is the ratio between the specific luminosity and the
specific binding energy ($L = \epsilon \times G M \dot{M}/R_{\rm NS}$,
$G$ is the gravitational constant, $M$ is the NS mass, $\dot{M}$ is
the accretion rate), and $L_{37}$ is the accretion luminosity in units
of $10^{37}$ erg/s, respectively. In this context we have $R_{\rm
disc} = r_{\rm m}$. Since the accretion disc is virialized, its
luminosity (averaged over the shot process) is $\langle I_{\rm DF}
\rangle = 0.5 \; \epsilon \; G\, M \dot{M} / r_{\rm m} $. The total
luminosity is $\langle I_{\rm DF} \rangle + \langle I_{\rm LC} \rangle
= \epsilon \; G\, M \dot{M} / R_{\rm NS} $. Therefore we have $\xi =
0.5 \; R_{\rm NS}/r_{\rm m} $. A lower limit for the inner rim of the
accretion disk is $r_{\rm m} \ge R_{\rm NS}$ and therefore we get the
constraint $\xi \leq 0.5$. This constraint is indicated as a dashed
horizontal line in fig. 3. This immediately gives a lower limit on the
fraction of coupled shots $\eta > 42\%$ indicated ad the dashed
vertical line in figures 2 and 3. This in turn limits the value of the
pulsed fraction below $0.37$ as indicated in fig. 2 by the dashed
horizontal line. An upper limit can be derived from the constraint
derived in the previous paragraph $\xi \geq 0.294$ that gives $r_{\rm
m} \le 1.7 \, R_{\rm NS}$. Adopting as a typical NS radius $R_{\rm NS}
\sim 10 $ km, we get $r_{\rm m} \leq 17$ km, that is consistent with
the requirement that accretion is not centrifugally
inhibited. Moreover adopting $\phi=0.3$ we can derive an upper limit
for the magnetic field of the NS. Adopting the explicit espression for
Alfv\'en radius in $r_{\rm m} \le 1.7 \, R_{\rm NS}$ we obtain $
\mu_{26} \leq 5.1 \; \phi_{0.3}^{-7/4} \; R_6^{9/4} m^{-1/4}
\epsilon^{-1/2} L_{37}^{1/2} $, where $phi_{0.3}$ is the $\phi$ factor
in units of $0.3$. Taking $m=1.4$, $\epsilon=1$ and $L_{37}=0.1 $ that
are appopriate for our source, we get $\mu_{26} \leq 1.5 \; R_6^{9/4}
$. This is well within the range $(1-5) \times 10^{26}$ G cm$^3$ that
are the lower and upper limits derived by Chakrabarty (199) and di
Salvo \& Burderi (2002) respectively.

This scenario implies that the aperiodic variability is generated in a
region affected by the lighthouse modulation, close to the neutron
star surface. Our result is in line with the coupling of the red noise
discovered analyzing the power spectra of some high mass X-ray
binaries (B97, LS97).  Our present result constrains some of the
models proposed for the origin of the aperiodic variability.  Indeed a
scenario like the one proposed by Aly \& Kuijpers (1990) in which the
aperiodic variability is generated by a shot noise process associated
with magnetospheric flares, caused by magnetic reconnection all around
the neutron star magnetosphere, cannot work for the origin of the
aperiodic variability in SAX J1808.4-3658 in case the red noise
component were coupled with the periodic modulation. On the other
hand, inhomogeneities could arise in the accretion flow during the
plasma penetration in the magnetosphere ({\it e.g.} via
Kelvin-Helmholtz instabilities, see B97 for a more extensive
discussion).  The subsequent impact of this inhomogenous accretion
flow funneled by the magnetic field lines onto the magnetic caps
appears to be a viable mechanism for the origin of the red noise.

\acknowledgments{  This  work was  partially supported  by a
grant   from  the   Italian  Ministry   of  University   and  Research
(Cofin 2001021123$\_$002). We warmly thank the referee  for very
useful and stimulating criticism.}

\clearpage

\begin{figure}
\epsscale{.45}      
\plotone{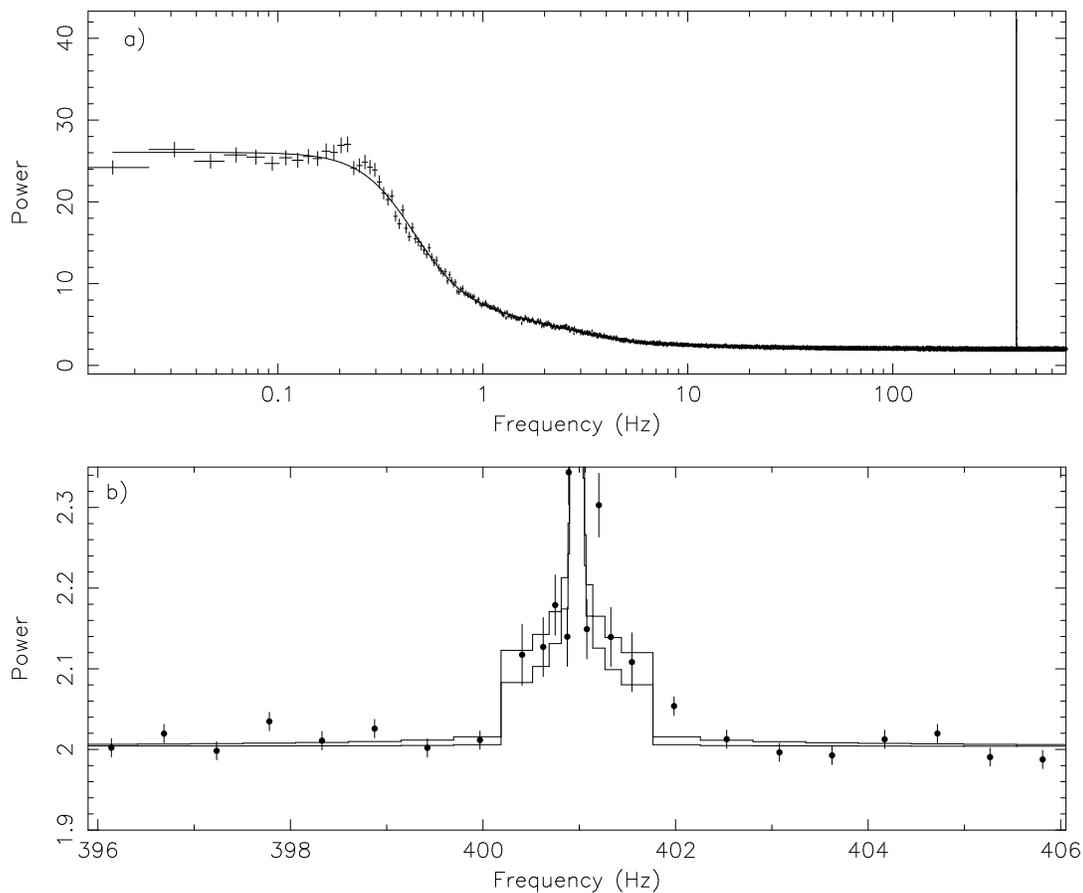}
\caption{{The power spectrum of SAX J1808.4-3658. Upper panel shows
    the measured broad band power spectrum and the simultaneous
    modelling of the red noise and harmonic components (solid line).
    The lower panel shows an expanded view of the region around the
    first fundamental harmonic: data have been binned by a factor 35
    to increase the signal to noise ratio. Both the fit with and
    without coupling with the red noise have been plotted. The lower
    line is without coupling. (Note that in the case of no coupling
    the residual broadening at the base of the harmonic line is due to
    the convolution of the delta function representing periodic
    modulation with the Fourier transform of the box function of
    duration T that multiplies the signal because of the finite
    duration of the observation).}
\label{fig:spe}}
\end{figure}

\clearpage

\begin{figure}
\plotone{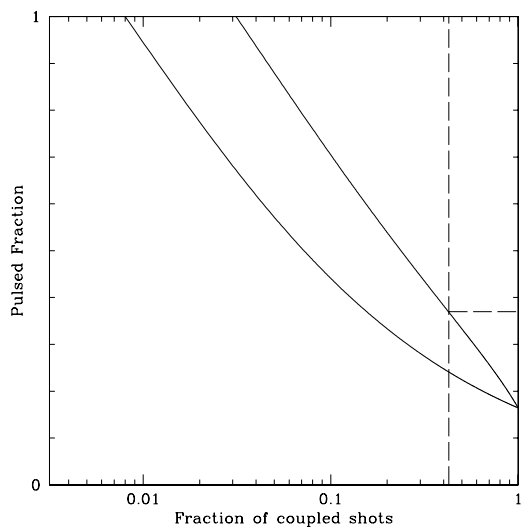}
\caption{{The ${\cal P_{\it f}}$ {\it vs} $\eta$ relation. The dashed
vertical line indicates a lower limit on the fraction of coupled shots
$\eta \ge 0.42$ derived from the constraints of Fig. 3. This, in turn,
poses an upper limit on the pulsed fraction $ {\cal P_{\it f}}\le 0.37$.}
\label{fig:pf}}
\end{figure}

\clearpage
\begin{figure}
\plotone{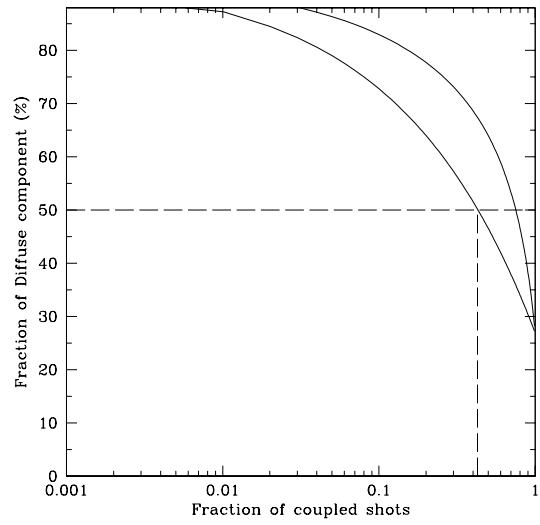}
\caption{{The $\xi$ {\it vs} $\eta$ relation. The horizontal line
indicates an upper limit for the fraction of diffuse component $\xi
\leq 50\%$. This, in turn, poses a lower limit on the fraction of
coupled shots $\eta \ge 0.42$, as shown by the dashed vertical line.}
\label{fig:xi}}
\end{figure}
\clearpage

\begin{table} 
\begin{center}
\caption{Observations log}
\begin{tabular}{crrc} 
\tableline \tableline
Observation ID   & Start Time        & End Time           & used in \\
                 &                   &                    & this work \\
\tableline 
 30411-01-01-00  &  11-4-1998  19:13 &  11-4-1998  21:56  & n \\
 30411-01-02-00  &  13-4-1998  01:23 &  13-4-1998  02:22  & n \\
 30411-01-03-00  &  16-4-1998  17:02 &  16-4-1998  22:59  & y \\
 30411-01-04-00  &  17-4-1998  01:11 &  17-4-1998  04:40  & y \\
 30411-01-05-00  &  18-4-1998  02:54 &  18-4-1998  09:17  & y \\
 30411-01-06-01  &  18-4-1998  12:27 &  18-4-1998  13:32  & y \\
 30411-01-06-00  &  18-4-1998  14:04 &  19-4-1998  01:02  & y \\
 30411-01-07-00  &  20-4-1998  20:52 &  20-4-1998  23:07  & y \\
 30411-01-08-00  &  23-4-1998  15:47 &  23-4-1998  23:19  & n \\
 30411-01-09-01  &  24-4-1998  15:51 &  24-4-1998  17:14  & n \\
 30411-01-09-02  &  24-4-1998  17:28 &  24-4-1998  23:21  & n \\
 30411-01-09-03  &  25-4-1998  14:17 &  25-4-1998  15:28  & n \\
 30411-01-09-04  &  25-4-1998  15:49 &  25-4-1998  21:40  & n \\
 30411-01-09-00  &  26-4-1998  15:55 &  26-4-1998  23:26  & n \\
 30411-01-10-02  &  27-4-1998  14:19 &  27-4-1998  15:41  & n \\
 30411-01-10-01  &  27-4-1998  15:57 &  27-4-1998  19:29  & n \\
 30411-01-10-00  &  29-4-1998  14:08 &  29-4-1998  18:56  & n \\
 30411-01-11-02  &  2-5-1998   01:14 &  2-5-1998   02:57  & n \\
 30411-01-10-03  &  2-5-1998   03:37 &  2-5-1998   04:38  & n \\
 30411-01-11-00  &  3-5-1998   18:59 &  3-5-1998   21:00  & n \\
 30411-01-11-01  &  6-5-1998   10:52 &  6-5-1998   14:00  & n \\

\tableline
\end{tabular}
\end{center}
\end{table}

\begin{table} 
\begin{center}
\caption{Results of the fit}
\medskip
\begin{tabular}{llll} 
\tableline \tableline
              & Parameter       & Value             & Error                                 \\
              &                 &                   & ($90\%$ confidence level)             \\
\tableline                                                                           
$WN$          &                 & $ 1.994$                   & $\pm 0.001$                  \\

\tableline                                                                           
$RN$          & $A_1$           & $  1.7\times        $      & $ \pm 6\times 10^{-1}  $     \\
              & $p_1$           & $  1.1\times 10^{1} $      & $ \pm 0.6\times 10^{1} $     \\
              & $p_2$           & $  2.6     $               & $ \pm 1 $                    \\
              & $p_3$           & $  1.6\times 10^{-1}$      & $\pm_9^8\times 10^{-2}  $    \\
              & $\tau_1$        & $  3.45\times 10^{-1}$     & $\pm_5^4\times 10^{-3} $     \\
              & $\alpha_1$      & $  3.3     $               & $\pm_2^3\times 10^{-1}  $    \\
              & $\tau_2$        & $  6.5\times 10^{-2} $     & $\pm_{0.5}^1\times 10^{-2}$  \\
              & $\alpha_2$      & $  2.0     $               & $ \pm 0.2 $                  \\
              & $\tau_3$        & $  3.9\times 10^{-3}  $    & $ \pm 1.\times 10^{-3} $     \\
              & $\alpha_3$      & $  1.5     $               & $ \pm 0.1 $                  \\
              & $A_2$           & $ 7.3\times 10^{1}  $      & $\pm 3$                      \\
              & $A_3$           & $  3.4\times 10^{-3}$      & $\pm 1\times 10^{-3} $       \\
\tableline
Coupling      & $r$             &    $0.002$                 & $\pm 0.001$                  \\

\tableline
\end{tabular}
\end{center}
\end{table}

\end{document}